\documentclass[12pt,aps,eqsecnum,nofootinbib,floatfix]{revtex4}

\usepackage{graphics}
\usepackage{graphicx}
\usepackage{epstopdf}
\usepackage{subfigure}
\usepackage{amssymb,amsmath}
\usepackage{CJK}
\usepackage{indentfirst}
\usepackage{amsmath}

\usepackage{varwidth}

\newcommand{\beq}[1]{\begin{eqnarray}\label{#1}}
\newcommand{\eeq}{\end{eqnarray}}

\begin{document}
			
\title{Entanglement entropy and monotones in scattering process}
	
\author{Jinbo Fan$^{1}$\footnote{email:~jinbofan@outlook.com},Gao-Ming Deng$^{2}$\footnote{email:~gmd2014cwnu@126.com},Xi-Jun Ren$^{1}$ \footnote{email:~997468504@qq.com}}
	
\medskip
	
\affiliation{\footnotesize $^1$ School of Physics and Electronics, Henan University, Kaifeng, 475004, China\\
		 $^2$ School of Physics and Space Science, China West Normal University, Nanchong 637002, China}
	
\begin{abstract}

In this paper, we study the entanglement property of a four-particle system. In this system, two initially entangled electrons A and C are scattered by two uncorrelated positrons B and D, respectively. We calculate the entanglements among the particles both before and after the double QED scattering ($AB\rightarrow AB, CD\rightarrow CD$). We find that the change of entanglement entropy between subsystems A and B during the scattering process is proportional to the total cross section, $\sigma_{tot}=\sigma_{AB}\times\sigma_{CD}$. Even though there is no direct interaction between subsystems A and C (or B and D), the scattering process induces entanglement change among them which is also proportional to $\sigma_{tot}$. This result shows some kind of entanglement sharing property in multipartite system. In order to further investigate the entanglement sharing, we calculate the entanglement monotones that quantify the genuine multipartite entanglement in a multipartite system. For our chosen scattering process, $e^+e^-\rightarrow\mu^+\mu^-$, however, we find that the outgoing state is a W-type four-partite entangled state that has no genuine 4-partite entanglement.

\end{abstract}

\maketitle 

\section{Introduction}

Numerous concepts in quantum information have received significant interest within the fundamental framework of quantum field theory, including entanglement entropy, error-correcting code, and complexity\cite{Takayanagi:2006,Pastawski:2015qua,Brown:2015lvg}. These issues provide some insight into the structure of interacting quantum field theories and spacetime geometry.

The physics of the scattering process plays a crucial role in various physical experiments and theories that probe the behavior of elementary particles. Nevertheless, previous theoretical investigations often focus on classical observables, such as the cross section and decay rate, as they are directly related to experiments. However, these classical properties cannot reveal all the information in the scattering process. For example, some soft particles can be produced during the scattering process which is not captured by classical observables, implying that some information is lost in scattering events. In the framework of quantum mechanics, entanglements between different degrees of freedom are generated in a scattering event, which provide us with a route to those with more detailed information.
 Furthermore, the entanglement entropy of scattering particles between high- and low-energy degrees of freedom in momentum space could also provide some deep insights into the scattering process\cite{Hsu:2012gk}. Balasubramanian\cite{Balasubramanian:2011wt} studied entanglement entropy of two divided momentum spaces with the perturbative calculations method, and this method was then followed by Refs.\cite{Park:2014hya,Peschanski:2016,Fan:2017hcd,Fan:2017mth,Sampaio:2016,Semenoff:2016,Ratzel:2016qhg} for the study of the entanglement in the 2-2 particles scattering process in a weakly coupled field theory. They found that the entanglement entropy changes during the scattering process; this variation in entanglement entropy from the initial to final state is proportional to the cross section. For a three-partite system, Sampaio \textit{et al.}\cite{Araujo:2019mni} considered a QED scattering (AB$\rightarrow$AB), wherein B is initially entangled with a third particle C that does not directly participate in the scattering process. They found that the spin expectation value $\langle\sigma_z\rangle$ of particle C does not change, but $\langle\sigma_x\rangle$ and $\langle\sigma_y\rangle$ are proportional to the total cross section of the AB scattering. 
 
 For a multiparticle system, the entanglement behavior of entangled states cannot be completely described by the entanglement entropy or mutual information. For example, the GHZ state $(\vert000\rangle+\vert111\rangle)/\sqrt{2}$ and W state $(\vert001\rangle+\vert010\rangle+\vert100\rangle)/\sqrt{3}$ are both entangled three-qubit state. When one party is traced out, the W state still exhibits two-partite entanglement, whereas the GHZ state does not exhibit any two-partite entanglement. In other words, they belong to different classes of three-qubit entangled states. For the pure states of bipartite systems, Bennett \textit{et al.}\cite{Bennett:1996gf} first discovered that partial entropy is a measure of entanglement for a given pure multipartite state. Subsequently, Wotters \textit{et al.}\cite{Hill:1997pfa,Wootters:1997id} evaluated three-partite pure states by exploiting the knowledge of mixed-state concurrence. Furthermore, Osterloh and Siewert \cite{Osterloh:2005} studied the entanglement monotones constructed with an antilinear operator filter, which can distinguish inequivalent classes of states with multipartite entanglement. Generally, the classification of entangled states is based on the applications of quantum information theory, rather than on specific mathematical forms. Two quantum states can complete the same quantum information task in a completely equivalent way, if they can be converted into each other under stochastic local operations and classical communication (SLOCC). As a result, the genuine multipartite entanglement in a multipartite system can be described and quantified by entanglement monotones under SLOCC. Accordingly, inequivalent classes of entangled states possess different entanglement monotones, and the entangled states with larger monotones can be transformed into entangled states with smaller monotones under SLOCC, but not vice versa. 

In this paper, we study the properties of entanglement among scattering particles ($AB\rightarrow AB, CD\rightarrow CD$), based on the double-scattering process $e^+e^-\rightarrow\mu^+\mu^-$, wherein electron $A$ and $C$ were initially entangled while positrons $B$ and $D$ were uncorrelated. This paper is organized as follows. In Sec. II, the variation of entanglement entropy between all two particles in the final state is calculated. In Sec. III,the entanglement monotones and classes of the final state are analyzed. Section IV presents the conclusions and final remarks.

\section{Particles entanglement of double scattering process}

For an elastic scattering process of two fermions in two-particle Fock space,  incoming and outgoing particle states  can be described as
\begin{align}
\vert p,s;q,r\rangle=\sqrt{2E_{\textbf{p}}}~{a^s_{\textbf{p}}}^{\dagger}\vert0\rangle_A\otimes \sqrt{2E_{\textbf{q}}}~{b^r_{\textbf{q}}}^{\dagger}\vert0\rangle_B,
\end{align}
where $\textbf{p}$ and $\textbf{q}$ are the 3-momenta of particles, and $s$, $r$ denote the spin of particles.
The fermionic creation/annihilation operators obey the anticommutation relations,
\begin{align}
\{a^s_{\textbf{p}}, {a^r_{\textbf{k}}}^\dagger\}=(2\pi)^3\delta^{(3)}(\textbf{p}-\textbf{k})\delta^{sr}
,~~~~
\{b^n_{\textbf{q}}, {b^m_{\textbf{\textit{l}}}}^\dagger\}=(2\pi)^3\delta^{(3)}(\textbf{q}-\textbf{\textit{l}})\delta^{nm}.
\end{align}
Then the inner product between two-particle states is defined as
\begin{align}
\langle k,s^\prime; l,r^\prime\vert p,s;q,r\rangle=2E_{\textbf{k}}2E_{\textbf{\textit{l}}}(2\pi)^3\delta^{(3)}(\textbf{k}-\textbf{p})  (2\pi)^3\delta^{(3)}(\textbf{\textit{l}}-\textbf{q})\delta^{ss^\prime}\delta^{rr^\prime}.
\end{align}

For a scattering process, the final state is determined by the initial state and the S matrix \cite{Park:2014hya},
\begin{align}
	\vert \textup{fin}\rangle=\int\frac{d^3\textbf{p}_3}{(2\pi)^32E_{\textbf{p}_3}}
	\int\frac{d^3\textbf{p}_4}{(2\pi)^32E_{\textbf{p}_4}}
	\sum_{r,s}\vert p_3,r;p_4,s\rangle\langle p_3,r;p_4,s\vert\textbf{S}\vert \textup{ini}\rangle.
\end{align}
The $\mathcal{T}$ matrix is defined as
\begin{align}
	&\langle p_3,r_3;p_4,r_4\vert i\mathcal{T}\vert p_1,r_1;p_2,r_2\rangle=(2\pi)^4\delta^{(4)}(p_1+p_2-p_3-p_4)\times i\mathcal{M}(r_1,r_2;r_3,r_4),
		\\ \notag
	&i\mathcal{T}=\textbf{S}-\textbf{1},
\end{align}
where shorthand notation $\mathcal{M}(r_1,r_2;r_3,r_4)$ is the invariant matrix element in a scattering process.

These concepts can be extended to a double scattering process, shown in the left of Fig. 1, $\mathcal{H}_{AB}:AB\rightarrow AB$ and $\mathcal{H}_{CD}:CD\rightarrow CD$, whose total Hilbert space is $\mathcal{H}_{\textup{tot}}=\mathcal{H}_{AB}\otimes\mathcal{H}_{CD}$. 
The concrete double scattering process we  consider here is shown in the right of Fig. 1, where a pair of initially entangled electrons ($AC$) collided with a pair of positrons ($BD$). Based on an exemplary QED process  $e^-e^+\rightarrow\mu^+\mu^-$, in the initial state, the subsystems $A, C$ correspond to two electrons $e^-$, and the subsystems $B, D$  two positrons $e^+$. In the final state, the subsystems $A, C$ correspond to two muons $\mu^-$, and the subsystems $B, D$ two antimuons $\mu^+$. We will study the behavior of entanglement in this four-particle system (ABCD) before and after double scattering.

\begin{figure}[htp]
  \centering
 \begin{varwidth}[htp]{\textwidth} 
  \vspace{0pt}
  \includegraphics[scale=0.32]{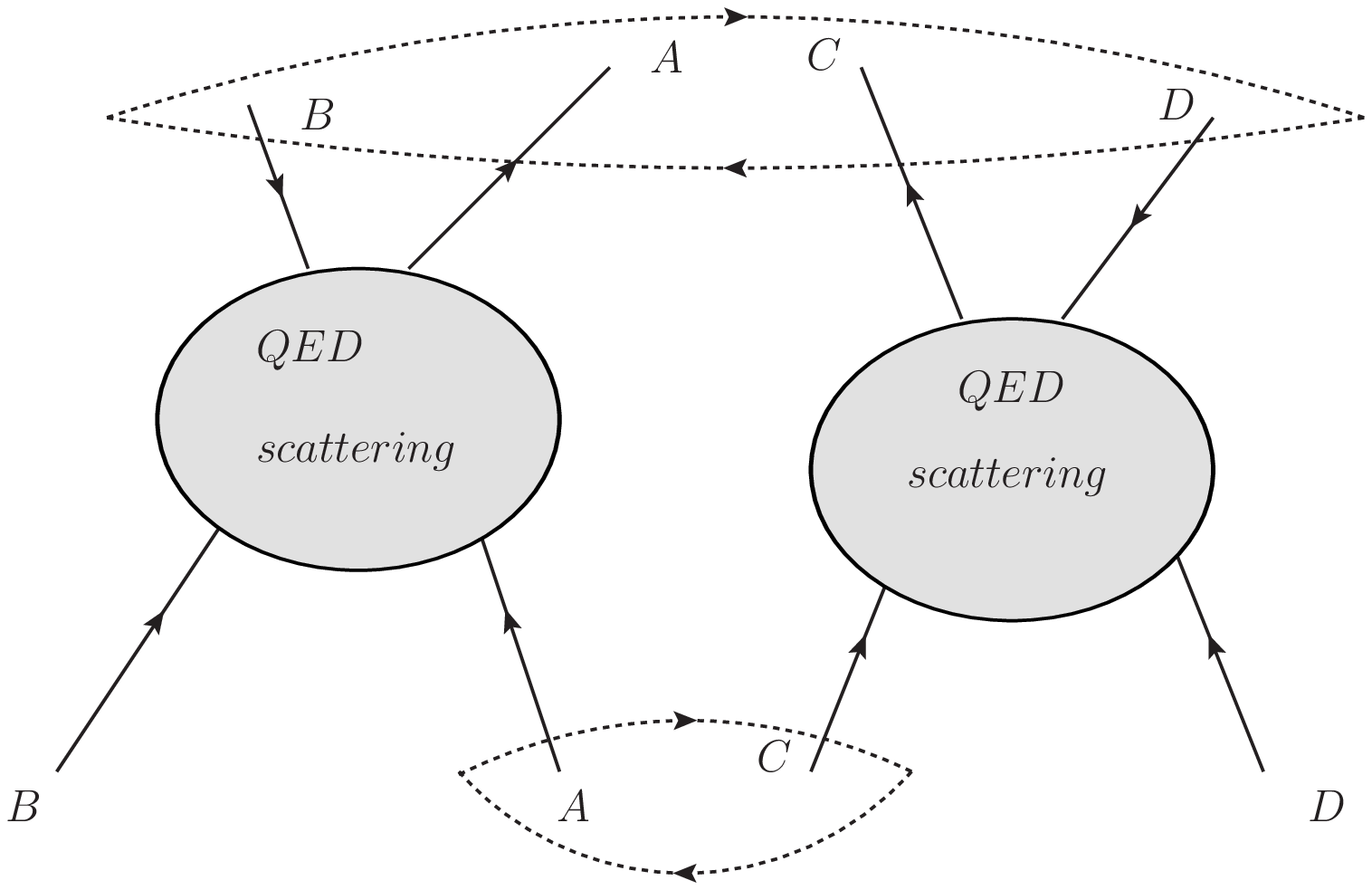}
  \end{varwidth}
  \qquad
  \qquad
  \begin{varwidth}[htp]{\textwidth}
  \vspace{0pt}
    \includegraphics[scale=0.36]{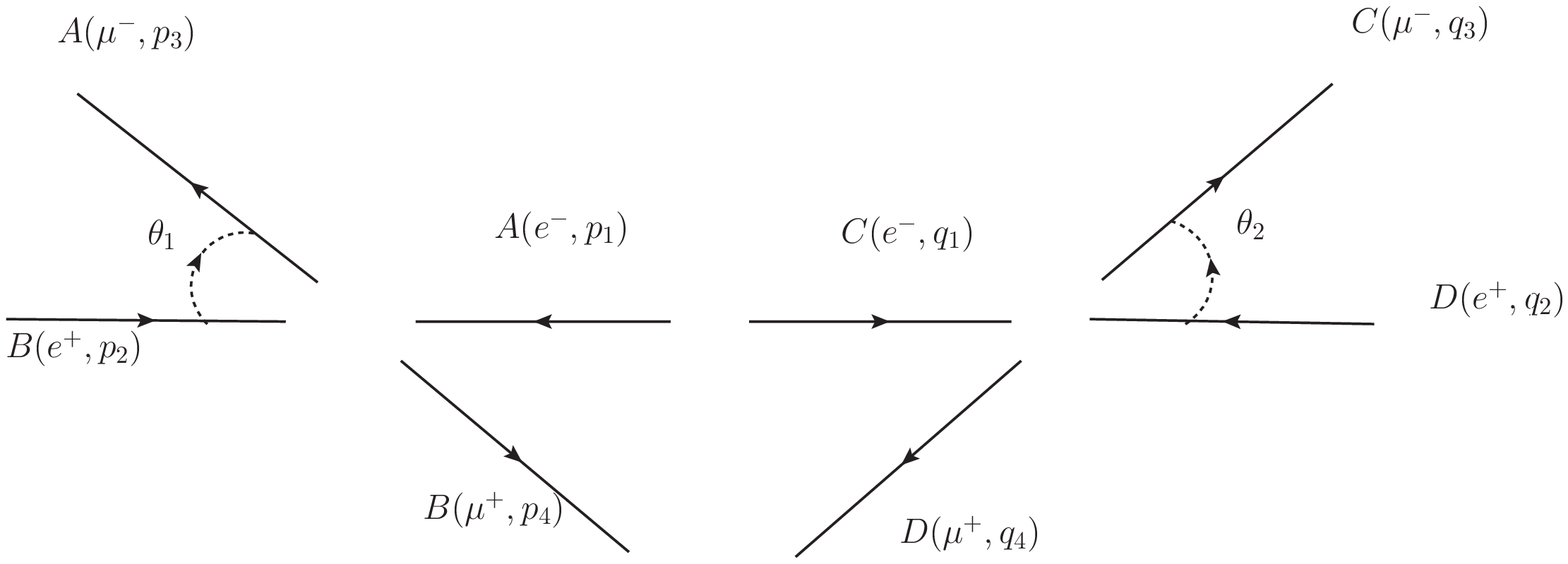}
 \end{varwidth}
 \caption{\textup{ Left: schematic illustration of the double scattering process. Right: the concrete scattering process based on two exemplary QED processes, $e^+e^-\rightarrow\mu^+\mu^-$, is considered in this paper. The process is shown in the center of mass frame.}}
\end{figure}

 The initial state is chosen as follows:
\begin{align}
\label{ini}
\vert\textup{ini}\rangle=\vert p_2,a\rangle\otimes
(\cos\eta\vert p_1,\uparrow; q_1,\uparrow\rangle+\sin\eta~ e^{i\beta}\vert p_1,\downarrow; q_1,\downarrow\rangle)
\otimes\vert q_2,b\rangle,
\end{align}
where $\eta\in[0,\pi/2]$ and $\beta\in[-\pi/2,3\pi/2]$ parametrizes the entanglement of the state.

The final state is given by
\begin{align}
\label{fin}	
\vert \textup{fin}\rangle=\vert \textup{ini}\rangle
-&\int_{\textbf{p}_3\neq\textbf{p}_1}d \Pi_2^{p_3}\int_{\textbf{q}_3\neq\textbf{q}_1}d \Pi_2^{q_3}
\biggr\{
\sum_{r_3,r_4}\sum_{s_3,s_4}
\biggr[\cos\eta \mathcal{M}_{AB}(\uparrow,a;r_3,r_4)\mathcal{M}_{CD}(\uparrow,b;s_3,s_4)
\\ \notag
&+e^{i\beta}\sin\eta\mathcal{M}_{AB}(\downarrow,a;r_3,r_4)\mathcal{M}_{CD}(\downarrow,b;s_3,s_4)\biggr]
\vert p_3,r_L;p_4,s_L\rangle
\otimes
\vert q_3,r_R;q_4,s_R\rangle
\biggr\},
\end{align}
where $\int d\Pi_2^{p_3}\equiv
\int\frac{d^3\textbf{p}_3}{(2\pi)^32E_{\textbf{p}_3}}
\int\frac{d^3\textbf{p}_4}{(2\pi)^32E_{\textbf{p}_4}}(2\pi)^4\delta^{(4)}(p_1+p_2-p_3-p_4)$.

 Our main task is to analyze the entanglement distribution in this state. In this section, we focus on the bipartite entanglement between any two scattering particles. The bipartite entanglement is measured by the mutual information. For example, 
\begin{align}
	I(A,C)=S(A)+S(C)-S(AC),
\end{align}
where $S(A), S(C)$ and $S(AC)$ are the von Neumann entropy of subsystems $A, C$ and $AC$.

To obtain an explicit formula, we choose to evaluate entanglement entropy in the center of mass frame, as shown in Fig. 1. The mass of electron is ignorable, 4-momentum of scattering particles can be written as 
\begin{align}
p_1&=(0,0,-E,E),~~~~p_2=(0,0,E,E),
~~~p_3=(-\sqrt{E^2-m^2}\sin\theta_1,0,-\sqrt{E^2-m^2}\cos\theta_1,E),
\\ \notag
p_4&=(\sqrt{E^2-m^2}\sin\theta_1,0,\sqrt{E^2-m^2}\cos\theta_1,E),
\\ \notag
q_1&=(0,0,E,E),~~~~q_2=(0,0,-E,E),
~~~q_3=(\sqrt{E^2-m^2}\sin\theta_2,0,\sqrt{E^2-m^2}\cos\theta_2,E),
\\ \notag
q_4&=(-\sqrt{E^2-m^2}\sin\theta_2,0,-\sqrt{E^2-m^2}\cos\theta_2,E),
\end{align}

At the lowest order, the scattering amplitude of process $e^+e^-\longrightarrow \mu^+\mu^-$ is
\begin{align}
\notag
\mathcal{M}(r_1,r_2;r_3,r_4)&\equiv\mathcal{M}(p_1,r_1;p_2,r_2\longrightarrow p_3,r_3;p_4,r_4)
\\ \notag
&=4\pi\alpha\bar{u}(p_3,r_3)[i\gamma^{\mu}]\nu(p_4,r_4)\frac{g_{\mu\nu}}{(p_1+p_2)^2}\bar{\nu}(p_2,r_2)[i\gamma^\nu] u(p_1,r_1),
\end{align}
where $\alpha=e^2/4\pi$ is the fine-structure constant, and mode functions $u(p,h)$ and $v(p,h)$ are chosen as helicity spinor in Ref. \cite{Thomson}, which helicity $h=(1/2,-1/2)$ denoted $(\uparrow,\downarrow)$. In the following calculations, the initial positron pair is chosen to be in the polarized state $a=b=\uparrow$.

\subsection{Subsystem AB}

For a general ($AB\rightarrow AB$) scattering process, it is found that the interaction will lead to the change of entanglement between the scattered particles and the change is proportional to the scattering cross section \cite{Park:2014hya,Fan:2017mth}. In the following, we will calculate the change of bipartite entanglement between the particles during our double scattering process ($AB\rightarrow AB, CD\rightarrow CD$). Obviously, for the given initial state (\ref{ini}), initial mutual information between subsystems A and B is zero. To get the mutual information between A and B for the final state, we will first determine the reduced density matrices, $\rho^f_{AB}, \rho^f_A, \rho^f_B$, then calculate their entropies, $S_{AB}^f, S^f_A, S^f_B$, which will lead us to their mutual information. 

By tracing out subsystem CD, we obtain the reduced density matrix of the subsystem AB,
\begin{align}
	\rho^f_{AB}=\frac{1}{\mathcal{N}}(\mathcal{IAB}+\mathcal{IIAB}).
\end{align}
The first term is,
\begin{align}
	\mathcal{IAB}=&2E_{\textbf{p}_1}2E_{\textbf{p}_2}2E_{\textbf{q}_1}2E_{\textbf{q}_2}V^4
	\biggr[
	\cos[\eta]^2\frac{\vert p_1,\uparrow;p_2,a\rangle\langle p_1,\uparrow;p_2,a\vert}
	{2E_{\textbf{p}_1}2E_{\textbf{p}_2}V^2}
	\\ \notag
	&+\sin[\eta]^2\frac{\vert p_1,\downarrow;p_2,a\rangle\langle p_1,\downarrow;p_2,a\vert}
	{2E_{\textbf{p}_1}2E_{\textbf{p}_2}V^2}
	\biggr],
\end{align}
The second term is,
\begin{align}
	\mathcal{IIAB}=&\alpha^4V^4
	2E_{\textbf{q}_1}2E_{\textbf{q}_2}2E_{\textbf{p}_1}2E_{\textbf{p}_2}
	\frac{T\int_{\textbf{p}_3\neq\textbf{p}_1}d \Pi_2^{p_3}}
	{2E_{\textbf{p}_1}2E_{\textbf{p}_2}V}
	\\ \notag
    &\times\biggr[
	\sum_{r_3,r^\prime_3}\sum_{r_4,r^\prime_4}
	\Lambda^{AB}(r_3,r^\prime_3;r_4,r^\prime_4)
	\frac{\vert p_3,r_3;p_4,r_4\rangle\langle p_3,r^\prime_3;p_4,r^\prime_4\vert}
	{2E_{\textbf{p}_3}2E_{\textbf{p}_4}V^2}
	\biggr].
\end{align}
where $\mathcal{N}$ is the normalization factor fixed by $tr_{AB}[\rho_{AB}^f]=1$. In performing partial traces, one finds Dirac deltas as $(2\pi)\delta^{(T)}(0)$ and $(2\pi)\delta^{(3)}(0)$, which come from background spacetime. The setting of the entire scattering process is designated to occur in a large spacetime volume of duration $T$ and spatial volume $V$; these factors are artifacts caused by regulating delta functions Ref.\cite{Weinberg},
\begin{align}
	\label{deltafnc}
	\delta^3_V(\textbf{p}-\textbf{p}^{\prime})=\frac{V}{(2\pi)^3}\delta_{\textbf{p},\textbf{p}^{\prime}},~~~~\delta_T(E_{\textup{if}})\equiv\delta_T(E_{\textup{fin}}-E_{\textup{ini}})=\frac{1}{2\pi}\int_{-T/2}^{T/2}dt~e^{i(E_{\textup{fin}}-E_{\textup{ini}})t},
\end{align}
which implies $V=(2\pi)^3\delta_V^{(3)}(0)$ and $T=(2\pi)\delta_T(0)$, respectively.

To simplify the expressions, we choose some shorthand notations,
\begin{align}
	&Ap(\downarrow,\uparrow;r_3,r_4;r^\prime_3,r^\prime_4)\equiv\frac{1}{\alpha^2}
	\mathcal{M}_{AB}(\downarrow,a;r_3,r_4)
	\mathcal{M}^\star_{AB}(\uparrow,a;r^\prime_3,r^\prime_4),
	\\
	&Aq(\downarrow,\uparrow;s_3,s_4;s^\prime_3,s^\prime_4)\equiv\frac{1}{\alpha^2}
	\mathcal{M}_{CD}(\downarrow,b; s_3,s_4)
	\mathcal{M}^\star_{CD}(\uparrow,b; s^\prime_3,s^\prime_4),
	\\
	&\mathcal{A}^{(q)}(\uparrow,\uparrow)=\frac{T\int_{\textbf{q}_3\neq\textbf{q}_1}d \Pi_2^{q_3}
		}{V2E_{\textbf{q}_1}2E_{\textbf{q}_2}}
		\sum_{s_3,s_4}Aq(\uparrow,\uparrow;s_3,s_4;s_3,s_4),
	\\ 
	&\Lambda^{(AB)}(r_3,r_4;r_3^\prime,r_4^\prime)
	\\ \notag
	=&\cos^2\eta~
    Ap(\uparrow,\uparrow;r_3,s_4;r_3^\prime,r_4^\prime)
	\mathcal{A}^{(q)}(\uparrow,\uparrow)
	+\sin^2\eta~
	Ap(\downarrow,\downarrow;r_3,r_4;r_3^\prime,r_4^\prime)
	\mathcal{A}^{(q)}(\downarrow,\downarrow)
	\\ \notag
	+&\cos\eta\sin\eta e^{-i\beta}~
	Ap(\uparrow,\downarrow;r_3,r_4;r_3^\prime,r_4^\prime)
	\mathcal{A}^{(q)}(\uparrow,\downarrow)
	+\cos\eta\sin\eta e^{i\beta}~
	Ap(\downarrow,\uparrow;r_3,r_4;r_3^\prime,r_4^\prime)
	\mathcal{A}^{(q)}(\downarrow,\uparrow),
	\\
	&
	\label{Lambda}
	\Lambda=
	\frac{T\int_{\textbf{p}_3\neq\textbf{p}_1}d \Pi_2^{p_3}}{V2E_{\textbf{p}_1}2E_{\textbf{p}_2}}
	\sum_{r_3,r_4}\Lambda^{(AB)}(r_3,r_4;r_3,r_4).
\end{align}
Then the normalization is given by
\begin{align}
	\label{normal}
	\mathcal{N}=2E_{\textbf{p}_1}2E_{\textbf{p}_2}2E_{\textbf{q}_1}2E_{\textbf{q}_2}V^4(1+\alpha^4\Lambda).
\end{align}

For the weak coupling theory, the reduced density matrix for subsystem AB at order $\alpha^4$ can be written as
\begin{align}
	\label{rAB}
	\rho_{AB}^{(f)}=diag((1-\alpha^4\Lambda)\mathcal{I}^{(AB)},\cdots,\alpha^4\Lambda^{(AB)}_{\theta_1},\cdots),
\end{align}
where the elements of this diagonal matrix correspond to \[\frac{\vert p_1,r_1;p_2,r_2\rangle\langle p_1,r^\prime_1;p_2,r^\prime_2\vert}
{2E_{\textbf{p}_1}2E_{\textbf{p}_2}V^2},\cdots, 	\frac{\vert p_3,r_3;p_4,r_4\rangle\langle p_3,r^\prime_3;p_4,r^\prime_4\vert}
{2E_{\textbf{p}_3}2E_{\textbf{p}_4}V^2},\]
the density matrix of a particular scattering angle $\theta_1$ in momentum space. For the unscattered direction $\theta_1=0$, the matrix
\[
\mathcal{I}^{(AB)}=
\begin{pmatrix}
	\cos[\eta]^2  &  0  & 0 & 0\\
	0  &  0  & 0 & 0\\
	0  &  0  & \sin[\eta]^2 & 0\\
	0  &  0  & 0 & 0\\
\end{pmatrix},
\]
has the eigenvalues $\{\cos[\eta]^2,\cos[\eta]^2,0,0\}$.
And for the scattered direction $\{\theta_1\neq0,\theta_2\neq0\}$, the matrix
\begin{small}
\begin{equation}
	\label{matrixAB}
	\Lambda^{(AB)}_{\theta_1}=
	\frac{1}{128\pi^2E^2V}
	\sqrt{\frac{E^2-m^2}{E^2}}
	\begin{pmatrix}
		\Lambda^{(AB)}(\uparrow\uparrow;\uparrow\uparrow) &  \Lambda^{(AB)}(\uparrow\uparrow;\uparrow\downarrow)  &\Lambda^{(AB)}(\uparrow\uparrow;\downarrow\uparrow)  &\Lambda^{(AB)}(\uparrow\uparrow;\downarrow\downarrow)  \\
		\Lambda^{(AB)}(\uparrow\downarrow;\uparrow\uparrow)  &  \Lambda^{(AB)}(\uparrow\downarrow;\uparrow\downarrow)  &\Lambda^{(AB)}(\uparrow\downarrow;\downarrow\uparrow)  &\Lambda^{(AB)}(\uparrow\downarrow;\downarrow\downarrow)  \\
		\Lambda^{(AB)}(\downarrow\uparrow;\uparrow\uparrow)  &  \Lambda^{(AB)}(\downarrow\uparrow;\uparrow\downarrow)  &\Lambda^{(AB)}(\downarrow\uparrow;\downarrow\uparrow)  &\Lambda^{(AB)}(\downarrow\uparrow;\downarrow\downarrow)  \\
		\Lambda^{(AB)}(\downarrow\downarrow;\uparrow\uparrow)  &  \Lambda^{(AB)}(\downarrow\downarrow;\uparrow\downarrow)  &\Lambda^{(AB)}(\downarrow\downarrow;\downarrow\uparrow)  &\Lambda^{(AB)}(\downarrow\downarrow;\downarrow\downarrow)
	\end{pmatrix},
\end{equation}
\end{small}
has the eigenvalues $\{R^{AB}_i=\frac{1}{128\pi^2E^2V}
\sqrt{\frac{E^2-m^2}{E^2}}a^{AB}_i,i=1,\dots,4\}$.

Then the entanglement entropy between subsystems AB and CD in the final state is
\begin{align}
	S^f_{AB}=&-(\cos[\eta]^2\log[\cos[\eta]^2]+\sin[\eta]^2\log[\sin[\eta]^2)
-\sum_i\int d\Omega_1
(\alpha^4R_i^{AB}\log[\alpha^4R_i^{AB}])
	\\ \notag
	&+\alpha^4\Lambda(1+\cos[\eta]^2\log[\cos[\eta]^2]+\sin[\eta]^2\log[\sin[\eta]^2])
+\mathcal{O}(\alpha^8).
\end{align}

 From density matrix $\rho_{AB}$ we can obtain reduced density matrix $\rho_{A}$ and $\rho_{B}$. 
 By tracing out the subsystem B, $\rho_A^f=Tr_B[\rho_{AB}^f]$, 
 the density matrix $\rho_A$ at $\alpha^4$ can be written as
 \begin{align}
 	\rho_{A}^{f}=diag((1-\alpha^4\Lambda)\mathcal{I}^{(A)},...,\alpha^4\Lambda^{(A)}_{\theta_1},...),
 \end{align}
 where the elements of this matrix correspond to  $\frac{\vert p_1,r_1\rangle\langle p_1,r^\prime_1\vert}
 {2E_{\textbf{p}_1}V},\cdots, 	\frac{\vert p_3,r_3\rangle\langle p_3,r^\prime_3\vert}
 {2E_{\textbf{p}_3}V}$. For the direction $\theta_1=0$, the matrix
 \[
 \mathcal{I}^{(A)}=
 \begin{pmatrix}
 	\cos^2\eta  &  0  \\
 	0  &  \sin^2\eta  \\
 \end{pmatrix},
 \]
 has the eigenvalues $\{\cos[\eta]^2,\sin[\eta]^2\}$. For the direction $\theta_1\neq0$, the matrix
 \begin{small}
 \begin{equation}
 	\label{matrixA}
 	\Lambda^{(A)}_{\theta_1}= 
 	\frac{T}{128\pi^2E^2V}\frac{\sqrt{E^2-m^2}}{E^2}
 	\begin{pmatrix}
 		\Lambda^{(A)}(\uparrow\uparrow)  &  \Lambda^{(A)}(\uparrow\downarrow)  \\
 		\Lambda^{(A)}(\downarrow\uparrow)  &  \Lambda^{(A)}(\downarrow\downarrow) \\
 	\end{pmatrix},
 ~~~~\Lambda^{(A)}(r_3,r^\prime_3)=\sum_{r_4}\Lambda^{(AB)}(r_3,r_4;r_3^\prime,r_4),
 \end{equation}
\end{small}
 has the eigenvalues $\{R_i^A, i=1,2\}$.The corresponding entanglement entropy for subsystem A is
 \begin{align}
 	\notag
 	S^f(A)
 	=&-(\cos[\eta]^2\log[\cos[\eta]^2 ]+\sin[\eta]^2\log[\sin[\eta]^2])-\sum_{i=1}\int d\Omega_1
 	(\alpha^4R^A_i\log[\alpha^4R^A_i])
 	\\
 	&+\alpha^4\Lambda(1+\cos[\eta]^2\log[\cos[\eta]^2 ]+\sin[\eta]^2\log[\sin[\eta]^2])
 	+\mathcal{O}(\alpha^8).
 \end{align}

By tracing out the subsystem A, $tr_A[\rho^f_{AB}]$, the density matrix $\rho_B$ at $\alpha^4$ can be written as{\Huge }
\begin{align}
	\rho_{B}^{f}=diag((1-\alpha^4\Lambda)\mathcal{I}^{(B)},...,\alpha^4\Lambda^{(B)}_{\theta_1},...),
\end{align}
where the elements of this matrix correspond  $\frac{\vert p_2,r_2\rangle\langle p_2,r^\prime_2\vert}
{2E_{\textbf{p}_2}V^2},\cdots, 	\frac{\vert p_4,r_4\rangle\langle p_4,r^\prime_4\vert}
{2E_{\textbf{p}_4}V^2}$. And the matrices
\[
\mathcal{I}^{(B)}=
\begin{pmatrix}
	1  &  0  \\
	0  &  0  \\
\end{pmatrix},
\]
\begin{equation}
	\label{matrixB}
 \Lambda^{(B)}_{\theta_1}= 
\frac{T}{128\pi^2E^2V}\frac{\sqrt{E^2-m^2}}{E^2}
\begin{pmatrix}
	\Lambda^{(B)}_{\uparrow\uparrow}  &  \Lambda^{(B)}_{\uparrow\downarrow}  \\
	\Lambda^{(B)}_{\downarrow\uparrow}  &  \Lambda^{(B)}_{\downarrow\downarrow} \\
\end{pmatrix},
 ~~~~\Lambda^{(B)}(r_4,r^\prime_4)=\sum_{r_3}\Lambda^{(AB)}(r_3,r_4;r_3,r^\prime_4),
\end{equation}
have the eigenvalues $\{1,0\}$ and $\{R_i^B, i=1,2\}$, respectively. The corresponding entanglement entropy of subsystem B is
\begin{align}
	S^f(B)=-\alpha^4\Lambda
	-\sum_{i=1}^{2}\int d\Omega_1
	(\alpha^4R^B_i\log[\alpha^4R^B_i])+\mathcal{O}(\alpha^8).
\end{align}

Thus, we obtain the variation of mutual information between subsystem A and B during the double scattering processes at leading order $\alpha^4\log\alpha^4$, 
\begin{align}
	\label{IAB}
	\triangle I(A,B)&=S^f(A)+S^f(B)-S^f(AB)-I^i(A,B)
	\\ \notag
	&=-\alpha^4\log[\alpha^4]
		\int d\Omega_1
	\biggr(
	\sum_{i=1}^{2}R_i^A
	+\sum_{j=1}^{2}R_j^B
	-\sum_{k=1}^{4}R_k^{AB}
	\biggr)
	+\mathcal{O}(\alpha^4),
\end{align}
where $\sum_{i=1}^{2}R_i^A, \sum_{i=1}^{2}R_i^B$ and $\sum_{i=1}^{4}R_i^{AB}$   are the ranks of matrices $(\ref{matrixA}), (\ref{matrixB})$ and $(\ref{matrixAB})$, respectively.

For the chosen initial state,
$
\vert\textup{ini}\rangle=\vert p_2,a\rangle\otimes
(\cos\eta\vert p_1,\uparrow; q_1,\uparrow\rangle+\sin\eta~ e^{i\beta}\vert p_1,\downarrow; q_1,\downarrow\rangle)
\otimes\vert q_2,b\rangle,
$
the total cross section at the lowest $\alpha^4$ for the double scattering $e^+e^-\rightarrow\mu^+\mu^-$ is
\begin{align}
	\sigma_{\vert ini\rangle}=\sigma_{AB}\times\sigma_{CD}=\begin{cases}
	\alpha^4\frac{(E^2-m^2)(2E^2+m^2)^2\pi^2}{9E^{10}\pi^2}\sin[\eta]^2,~~~~~~~~ a=b=\uparrow  \\
		\alpha^4\frac{(E^2-m^2)(2E^2+m^2)^2\pi^2}{9E^{10}\pi^2}\cos[\eta]^2,~~~~~~~~ a=b=\downarrow 
	\end{cases}
\end{align}
Finally, for the initial polarized system BD, detailed calculations show that the variation of mutual information between subsystems A and B at leading order is
\begin{align}
	\label{tIAB}
	\triangle I(A,B)
	&=-\alpha^4\log[\alpha^4]\frac{T^2}{V^2}f_{AB}(E_{cm},m,\eta)+\mathcal{O}(\alpha^4)
	\\ \notag
	&=-4\log[\alpha^4]\frac{T^2\sigma_{\vert ini\rangle}}{V^2}+\mathcal{O}(\alpha^4).
\end{align}
where $E_{cm}=4E$ stands for the total energy in four-particle system. The functions  $f_{AB}$ appearing in the mutual information is plotted in Fig. 2. Obviously, it is proportional to the scattering cross section $\sigma_{\vert ini\rangle}$

\begin{figure}[htp]
	\centering
	\begin{varwidth}[htp]{\textwidth} 
		\vspace{0pt}
		\includegraphics[scale=0.5]{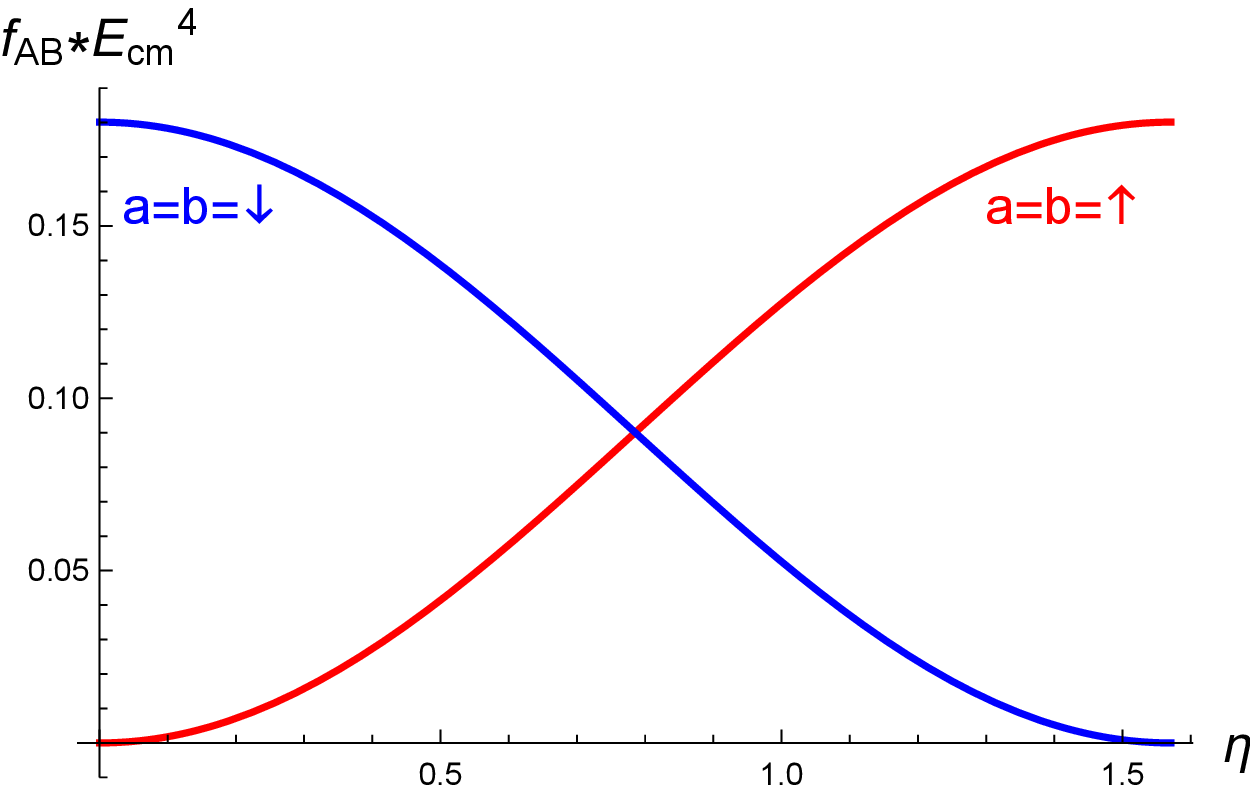}
	\end{varwidth}
  \qquad
\qquad
\begin{varwidth}[htp]{\textwidth}
	\vspace{0pt}
	\includegraphics[scale=0.55]{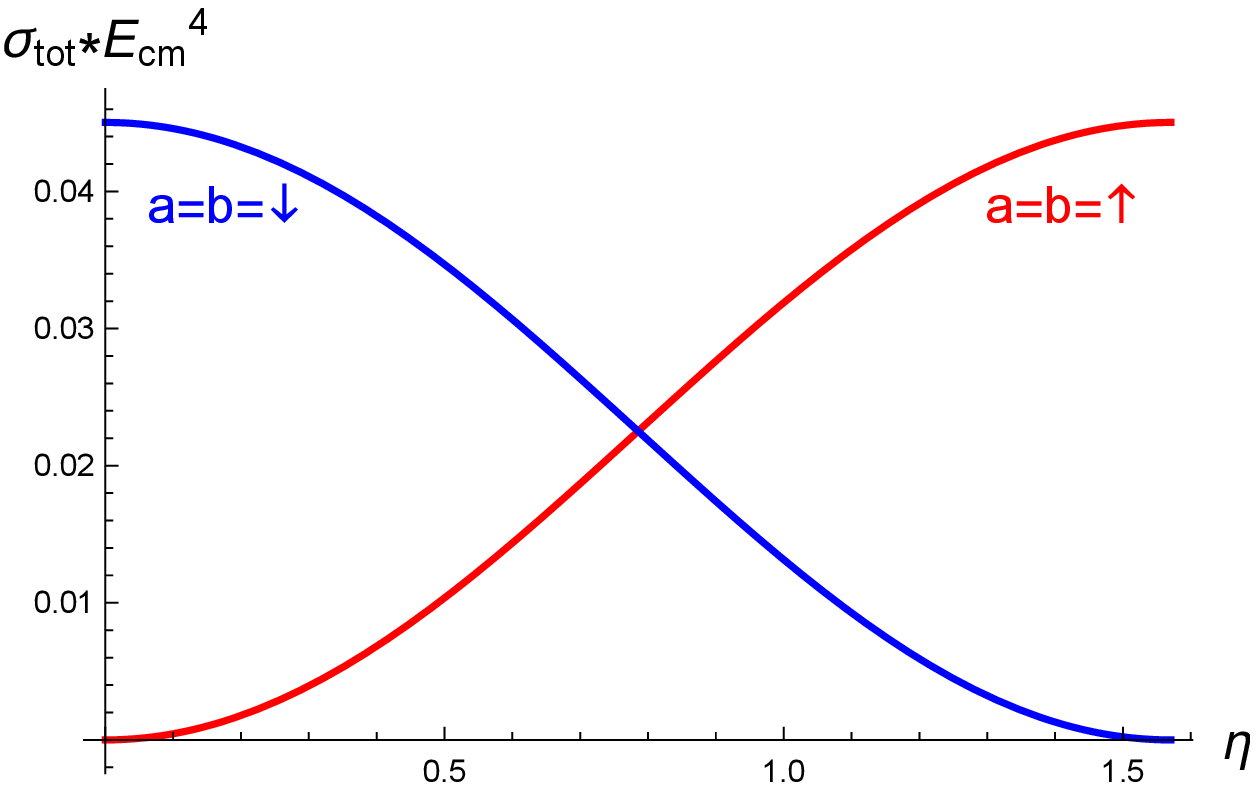}
\end{varwidth}
	\caption{\textup{The $\alpha^4\log\alpha^4$ order contributions to variation of mutual information $I(A,B)$ and the total cross section $\sigma_{\vert ini\rangle}$ as a function of the entanglement parameter $\eta$ of the initial state.}}
\end{figure}

Equation $(\ref{tIAB})$ and Fig. 2 show that, the change of entanglement between subsystems A and B during the double scattering process is proportional to the total cross section $\sigma_{AB}\times\sigma_{CD}$ rather than $\sigma_{AB}$ alone. Therefore, the scattering process ($CD\rightarrow CD$) affects the entanglement between A and B, given the initial entanglement between subsystems A and C. This illustrates the global sharing property of entanglement in a multipartite system.

 In fact, the function $f_{AB}$ is related to the quantity $\Lambda$ in a normalization factor ($\ref{normal}$) as
\begin{align}
	f_{AB}=\frac{V^2}{T^2}\Lambda=\frac{4}{\alpha^4}\sigma_{\vert ini\rangle}.
\end{align}
As shown in Ref. \cite{Araujo:2019mni}, if the subsystems A and B experience other QED processes, like Bhabha scattering $e^-e^+\to e^-e^+$,Møller scattering $e^-e^-\to e^-e^-$ and a process $e^-\mu^-\to e^-\mu^-$, the cross section is diverging because of collinearity. Then the factor $\Lambda\to \infty$ and function $f_{AB}$ will also diverge. For elimination of divergence, the contributions of soft photon scattering should be included. However, when soft photons are included, the four-particle final state will become a mixed state after tracing out these soft photons which leads to difficulty in calculations.

As a compromise, we may consider the change of entanglement entropy along a particular scattering angle $\theta_1$ in momentum space,
\begin{align}
	\triangle I_{\theta_1}(A,B)&=S_{\theta_1}^f(A)+S_{\theta_1}^f(B)-S_{\theta_1}^f(AB)
	\\ \notag
	&=-\alpha^4\log[\alpha^4]
	\biggr(
	\sum_{i=1}^{2}R_i^A
	+\sum_{j=1}^{2}R_j^C
	-\sum_{k=1}^{4}R_k^{AC}
	\biggr)
	+\mathcal{O}(\alpha^4)
	\\ \notag
	&=-4\log[\alpha^4]\frac{T^2}{V^2}\frac{d\sigma_{tot}}{d\Omega_1}	+\mathcal{O}(\alpha^4),
\end{align}
where $d\sigma_{tot}/d\Omega_1$ is the differential cross section for the scattering rate into an element of solid angle $d\Omega_1=d(\cos\theta_1)d\phi$. Since the change of entanglement entropy is proportional to the differential cross section, it is reasonable to expect that, for other QED scattering process, the change of entanglement entropy is also proportional to the total cross section.

\subsection{Subsystems AC and BD}

During the double scattering process, subsystems A and C (also B and D) have no direct interactions, but as we will show in this section their entanglement or mutual information also change. Their initial mutual information can be easily calculated with the initial state $(\ref{ini})$, where $I^i(AC)=-2[\cos^2\eta\ln[\cos^2\eta]+\sin^2\eta\ln[\sin^2\eta]]$.

By tracing out the subsystem BD of the final state $(\ref{fin})$, $\rho^f_{AC}=\frac{1}{\mathcal{N}}Tr_{(BD)}{[\rho^f]}$, we obtain the reduced density matrix of the subsystem AC,
\begin{align}
	\rho^f_{AC}=\frac{1}{\mathcal{N}}(\mathcal{IAC}+\mathcal{IIAC}).
\end{align}
The first term is
\begin{align}
	\mathcal{IAC}&=2E_{\textbf{p}_1}2E_{\textbf{p}_2}2E_{\textbf{q}_1}2E_{\textbf{q}_2}V^4
	\biggr[\cos^2\eta\frac{\vert p_1,\uparrow;q_1,\uparrow\rangle\langle p_1,\uparrow;q_1,\uparrow\vert}{2E_{\textbf{p}_1}2E_{\textbf{q}_1}V^2}
	+\sin^2\eta\frac{\vert p_1,\downarrow;q_1,\downarrow\rangle\langle p_1,\downarrow;q_1,\downarrow\vert}
	{2E_{\textbf{p}_1}2E_{\textbf{q}_1}V^2}
	\\ \notag
	&+\sin\eta\cos\eta e^{-i\beta}\frac{\vert p_1,\uparrow;q_1,\uparrow\rangle\langle p_1,\downarrow;q_1,\downarrow\vert}
	{2E_{\textbf{p}_1}2E_{\textbf{q}_1}V^2}
	+\sin\eta\cos\eta e^{i\beta}\frac{\vert p_1,\downarrow;q_1,\downarrow\rangle\langle p_1,\uparrow;q_1,\uparrow\vert}
	{2E_{\textbf{p}_1}2E_{\textbf{q}_1}V^2}
	\biggr].
\end{align}
The second term is
\begin{align}
	&\mathcal{IIAC}
	\\ \notag
	=&\alpha^4T^2V^2
	\int_{\textbf{p}_3\neq\textbf{p}_1}d \Pi_2^{p_3}
	\int_{\textbf{q}_3\neq\textbf{p}_1}d \Pi_2^{q_3}
	\biggr[
	\sum_{r_3,r^\prime_3}\sum_{s_3,s^\prime_3}
	\Lambda^{(AC)}(r_3,s_3;r^\prime_3,s^\prime_3)
	\frac{\vert p_3,r_3;q_3,s_3\rangle\langle p_3,r^\prime_3;q_3,s^\prime_3\vert}
	{2E_{\textbf{p}_3}2E_{\textbf{q}_3}V^2}
	\biggr],
\end{align}
Choosing a shorthand notation,
\begin{align}
&\Lambda^{(AC)}(r_3,s_3;r_3^\prime,s_3^\prime)
=\sum_{r_4,s_4}
\cos^2\eta~
Ap(\uparrow,\uparrow;r_3,r_4;r_3^\prime,r_4)
Aq(\uparrow,\uparrow;s_3,s_4;s_3^\prime,s_4)
\\ \notag
&+\sin^2\eta~
Ap(\downarrow,\downarrow;r_3,r_4;r_3^\prime,r_4L)
Aq(\downarrow,\downarrow;s_3,s_4;s_3^\prime,s_4)
\\ \notag
&+\cos\eta\sin\eta e^{-i\beta}~
Ap(\uparrow,\downarrow;r_3,r_4;r_3^\prime,r_4)
Aq(\uparrow,\downarrow;s_3,s_4;s_3^\prime,s_4)
\\ \notag
&+\cos\eta\sin\eta e^{i\beta}~
Ap(\downarrow,\uparrow;r_3,r_4;r_3^\prime,r_4)
Aq(\downarrow,\uparrow;s_3,s_4;s_3^\prime,s_4),
\end{align}
It can be easily shown that the factor $\Lambda$ in Eq. ($\ref{Lambda}$) satisfies
\begin{align}
	\Lambda=
	\frac{T\int_{\textbf{p}_3\neq\textbf{p}_1}d \Pi_2^{p_3}}{V2E_{\textbf{p}_1}2E_{\textbf{p}_2}}
	\frac{T\int_{\textbf{q}_3\neq\textbf{q}_1}d \Pi_2^{q_3}}{V2E_{\textbf{q}_1}2E_{\textbf{q}_2}}
	\sum_{r_3,s_3}\Lambda^{(AC)}(r_3,s_3;r_3,s_3).
\end{align}
The reduced density matrix for subsystem AC at order $\alpha^4$ can be written as
\begin{align}
	\rho_{AC}^{(f)}=diag((1-\alpha^4\Lambda)\mathcal{I}^{(AC)},...,\alpha^4\Lambda^{(AC)}_{\theta_1,\theta_2},...),
\end{align}
where the elements of this matrix correspond to \[ \frac{\vert p_1,r_1;q_1,s_1\rangle\langle p_1,r^\prime_1;q_1,s^\prime_1\vert}
{2E_{\textbf{p}_1}2E_{\textbf{q}_1}V^2}, \cdots, \frac{\vert p_3,r_3;q_3,s_3\rangle\langle p_3,r^\prime_3;q_3,s^\prime_3\vert}
{2E_{\textbf{p}_3}2E_{\textbf{q}_3}V^2},\cdots,\] the density matrix of scattering angle $(\theta_1,\theta_2)$ in momentum space.
For the direction $\theta_1=0,\theta_2=0$, the matrix
\[
\mathcal{I}^{(AC)}=
\begin{pmatrix}
	\cos^2\eta  &  0  & 0 & \sin\eta\cos\eta e^{-i\beta}\\
	0  &  0  & 0 & 0\\
	0  &  0  & 0 & 0\\
	\sin\eta\cos\eta e^{i\beta}  &  0  & 0 & \sin^2\eta\\
\end{pmatrix},
\]
has the eigenvalues $\{0,0,0,1\}$. For the direction $\theta_1\neq0,\theta_2\neq0$, the matrix

\begin{small}
\begin{equation}
	\label{matrixAC}
	\Lambda^{(AC)}_{\theta_1,\theta_2}=
	(\frac{T}{128\pi^2E^2V})^2(\frac{E^2-m^2}{E^2})
	\begin{pmatrix}
		\Lambda^{(AC)}(\uparrow\uparrow;\uparrow\uparrow) &  \Lambda^{(AC)}(\uparrow\uparrow;\uparrow\downarrow)  &\Lambda^{(AC)}(\uparrow\uparrow;\downarrow\uparrow)  &\Lambda^{(AC)}(\uparrow\uparrow;\downarrow\downarrow)  \\
		\Lambda^{(AC)}(\uparrow\downarrow;\uparrow\uparrow)  &  \Lambda^{(AC)}(\uparrow\downarrow;\uparrow\downarrow)  &\Lambda^{(AC)}(\uparrow\downarrow;\downarrow\uparrow)  &\Lambda^{(AC)}(\uparrow\downarrow;\downarrow\downarrow)  \\
		\Lambda^{(AC)}(\downarrow\uparrow;\uparrow\uparrow)  &  \Lambda^{(AC)}(\downarrow\uparrow;\uparrow\downarrow)  &\Lambda^{(AC)}(\downarrow\uparrow;\downarrow\uparrow)  &\Lambda^{(AC)}(\downarrow\uparrow;\downarrow\downarrow)  \\
		\Lambda^{(AC)}(\downarrow\downarrow;\uparrow\uparrow)  &  \Lambda^{(AC)}(\downarrow\downarrow;\uparrow\downarrow)  &\Lambda^{(AC)}(\downarrow\downarrow;\downarrow\uparrow)  &\Lambda^{(AC)}(\downarrow\downarrow;\downarrow\downarrow)
	\end{pmatrix},
\end{equation}
\end{small}
has the eigenvalues $\{R^{AC}_i=(\frac{T}{128\pi^2E^2V})^2(\frac{E^2-m^2}{E^2})a^{AC}_i,i=1...4\}$. The corresponding entanglement entropy of subsystem AC is
\begin{align}
	S^f_{AC}=\alpha^4\Lambda-\sum_i\int d\Omega_1\int d\Omega_2
	(\alpha^4R_i^{AC}\log[\alpha^4R_i^{AC}])
	+\mathcal{O}(\alpha^8).
\end{align}
Based on the symmetry of the system (ABCD), we have $S(A)=S(C)$. Thus, we obtain the variation of mutual information between subsystems A and C at leading order,
\begin{align}
	\label{IAC}
	\triangle I(A,C)
	&=-\alpha^4\log[\alpha^4]\frac{T^2}{V^2}f_{AC}(E_{cm},m,\eta)+\mathcal{O}(\alpha^4)
	\\ \notag
	&=-4\log[\alpha^4]\frac{T^2\sigma_{\vert ini\rangle}}{V^2}+\mathcal{O}(\alpha^4).
\end{align}

For the initial polarized positron pair (BD), the final state is pure, and the corresponding entanglement entropy has $S^f(AC)=S^f(BD)$. The particles B and D of the final state obey the transformation $\theta_1\rightarrow\pi-\theta_2$, and one can obtain $S^f(B)=S^f(D)$.
Thus, the variation of mutual information between subsystems B and D at leading order is
\begin{align}
	\triangle I(B,D)&=-\alpha^4\log[\alpha^4]\frac{T^2}{V^2}f_{BD}(E_{cm},m,\eta)+\mathcal{O}(\alpha^4) \\ \notag
	&=-4\log[\alpha^4]\frac{T^2\sigma_{\vert ini\rangle}}{V^2}+\mathcal{O}(\alpha^4).
\end{align}
The functions  $f_{AC}$ and $f_{BD}$ appearing in the entanglement entropy are plotted in Fig. 3.

\begin{figure}[htp]
	\centering
	\begin{varwidth}[htp]{\textwidth} 
		\vspace{0pt}
		\includegraphics[scale=0.5]{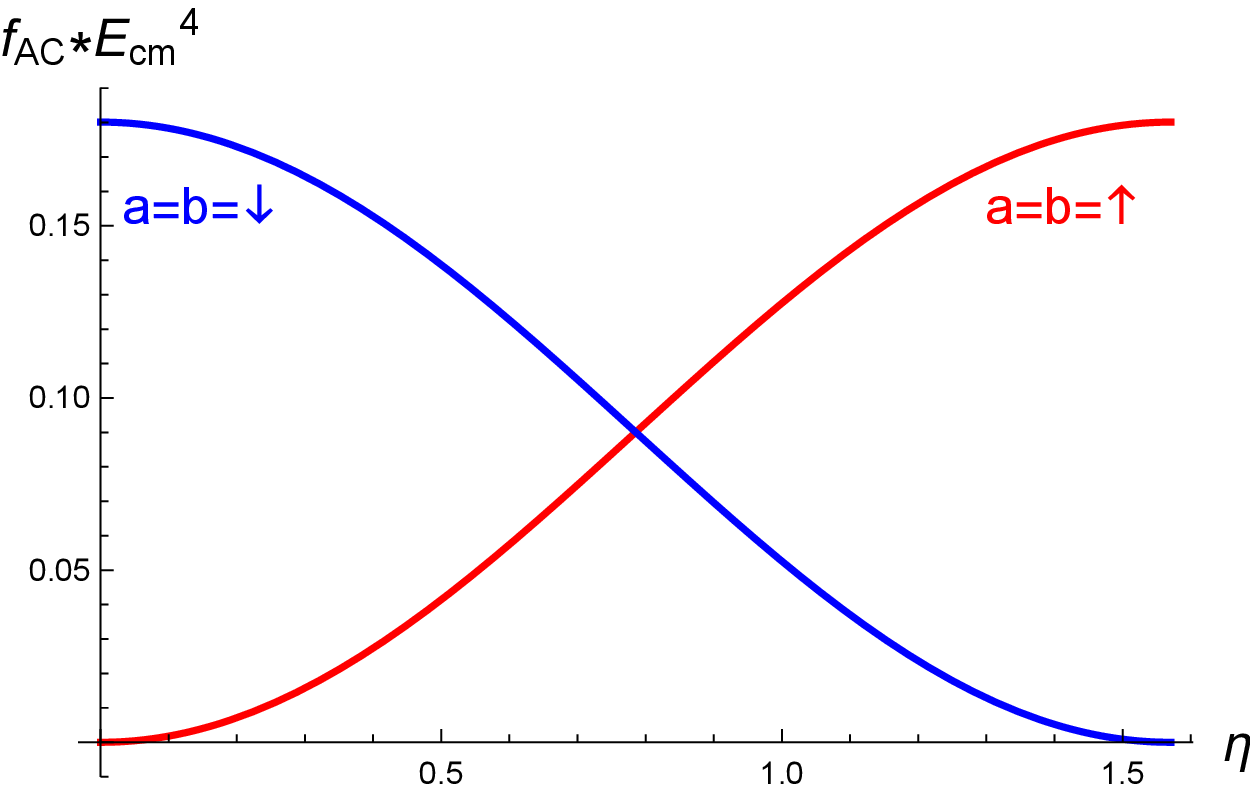}
	\end{varwidth}
	\qquad
	\qquad
	\begin{varwidth}[htp]{\textwidth} 
		\vspace{0pt}
		\includegraphics[scale=0.5]{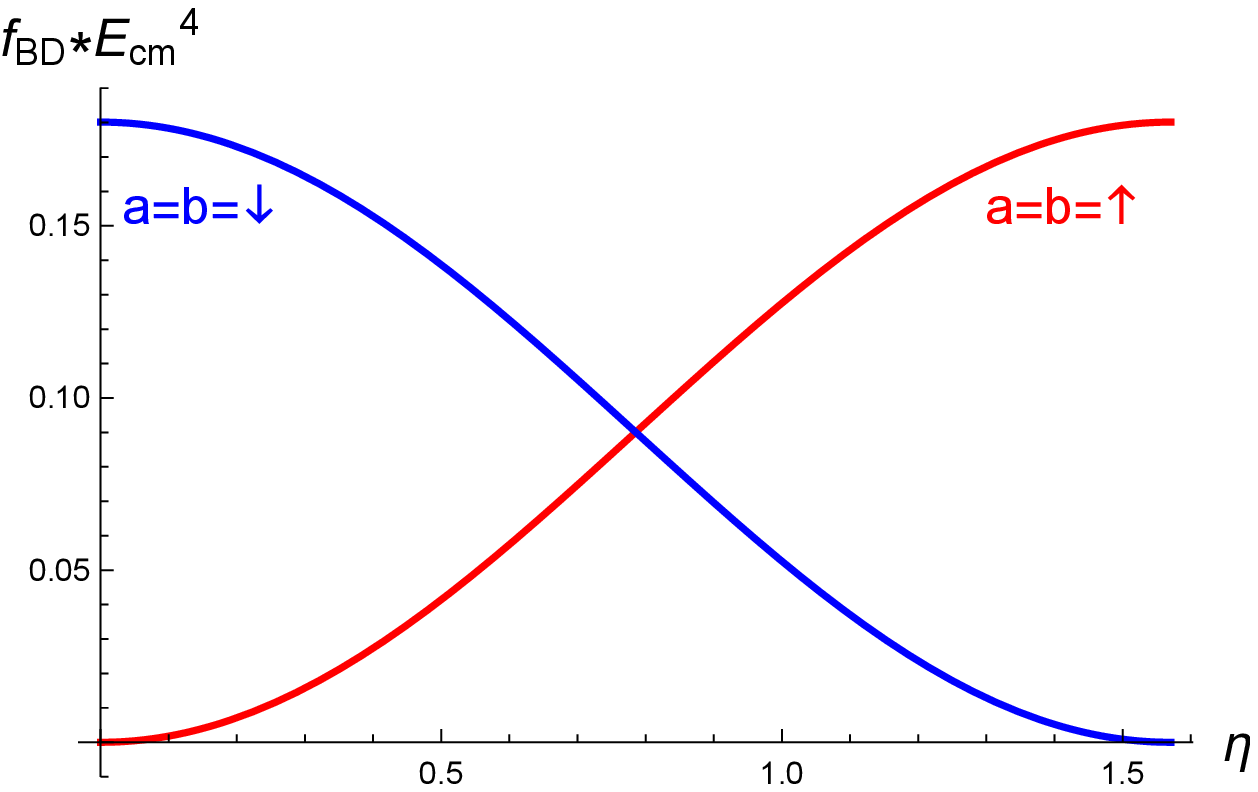}
	\end{varwidth}
	\caption{\textup{The $\alpha^4\log\alpha^4$ order contributions to mutual information change among scattering particles is a function of the entanglement parameter $\eta$ of the intital state. The left figure is  for subsystem AC. 
			The right figure is for subsystem BD.}}
\end{figure}

If subsystems A and C are initially untangled, then the double scattering process reduces completely into two separate scattering processes, and the final state will be $\vert \textup{fin}\rangle=\vert \textup{fin}\rangle_{AB}\otimes\vert \textup{fin}\rangle_{CD}$, where no entanglement exist among subsystems A and C or B and D. However, if subsystems A and C are initially entangled, even though they do not have direct interactions during the scattering process, their entanglement still changes proportionally to the total cross section of the double scattering  process. Our calculations show that the bipartite entanglement distributes among all pairs of subsystems after the scattering process  ($AB\rightarrow AB, CD\rightarrow CD$).

\section{Monotones of scattering particles}

From the above calculations, for the double-scattering process $e^+e^-\rightarrow\mu^+\mu^-$, entanglement is present among the final particles. Reference. \cite{Verstraete:2002} shows that there exist nine families of states corresponding to nine different ways of entangling four qubits. This classification method is based on the calculation of the entanglement monotones of entangled states, which can distinguish inequivalent types of genuine four-qubit entanglement. The entanglement monotone for a given state is zero if the state is separable, i.e., a one- or two-qubit part exists that can be factored out. Identification of the families of the scattering final state, can provide a way to prepare specific genuine multiparticles entangled states.

The three entanglement monotones, named filters in Ref.\cite{Osterloh:2006}, are constructed with antilinear operators as follows:
\begin{align}
\mathcal{F}^{(4)}_1&=(\sigma_{\mu}\sigma_{\nu}\sigma_y\sigma_y)\cdot(\sigma^{\mu}\sigma_y\sigma_{\lambda}\sigma_y)\cdot(\sigma_y\sigma^{\nu}\sigma^{\lambda}\sigma_y),
\\
\mathcal{F}^{(4)}_2&=(\sigma_{\mu}\sigma_{\nu}\sigma_y\sigma_y)\cdot(\sigma^{\mu}\sigma_y\sigma_{\lambda}\sigma_y)\cdot(\sigma_y\sigma^{\nu}\sigma_y\sigma_\tau)\cdot(\sigma_y\sigma_y\sigma_{\lambda}\sigma^\tau),
\\
\mathcal{F}^{(4)}_3&=\frac{1}{2}(\sigma_{\mu}\sigma_{\nu}\sigma_y\sigma_y)\cdot(\sigma^{\mu}\sigma^{\nu}\sigma_y\sigma_y)\cdot(\sigma_\rho\sigma_y\sigma_\tau\sigma_y)\cdot(\sigma_\rho\sigma_y\sigma_{\tau}\sigma_y)\cdot(\sigma_y\sigma_\rho\sigma_\tau\sigma_y)\cdot(\sigma^\rho\sigma_y\sigma^\tau\sigma_y),
\end{align}
where  $g^{\mu\nu}=diag\{-1,1,0,1\}$ and Pauli matrices $\sigma_0:=1, \sigma_1:=\sigma_x, \sigma_2=\sigma_y, \sigma_3:=\sigma_z$.
Next, we calculate the entanglement monotones of the final state.

 Here we are interested in the final state of the double scattering process,
\begin{align}
\vert \textup{fin}\rangle=\vert \textup{ini}\rangle
-&\int_{\textbf{p}_3\neq\textbf{p}_1}d \Pi_2^{p_3}\int_{\textbf{q}_3\neq\textbf{q}_1}d \Pi_2^{q_3}
\biggr\{
\sum_{r_3,r_4}\sum_{s_3,s_4}
\biggr[\cos\eta \mathcal{M}_{AB}(\uparrow,a;r_3,r_4)\mathcal{M}_{CD}(\uparrow,b;s_3,s_4)
\\ \notag
&+e^{i\beta}\sin\eta\mathcal{M}_{AB}(\downarrow,a;r_3,r_4)\mathcal{M}_{CD}(\downarrow,b;s_3,s_4)\biggr]
\vert p_3,r_3;p_4,r_4\rangle
\otimes
\vert q_3,s_3;q_4,s_4\rangle
\biggr\},
\end{align}
where $a$ and $b$ are chosen as $\uparrow$ or $\downarrow$, because the condition for calculating monotones is for the pure state.

Through direct calculations, we find that for both $a=b=\uparrow$ and $a=b=\downarrow$ cases, the entanglement monotones of this final state are zero,
\begin{align}
	\langle\textup{fin}^{\star}\vert\mathcal{F}_1^{(4)}\vert\textup{fin}\rangle
	=0,~~~~\langle \textup{fin}^{\star}\vert\mathcal{F}_2^{(4)}\vert\textup{fin}\rangle&=0,~~~~\langle \textup{fin}^{\star}\vert\mathcal{F}_3^{(4)}\vert \textup{fin}\rangle=0.
\end{align}

As mentioned above, when an entangled electron pair collides with a polarized positron beam via double-scattering process $e^+e^-\rightarrow\mu^+\mu^-$, entanglement are present among random scattering particles. However, entanglement monotones for the final state show that the scattering final state is not a genuine mutiparticles entangled state, and that the state of the four entangled particles  is similar to the W state. In other words, if the entanglement of two particles ($AB$) in the final state is broken, other particles ($AC$) still remain entangled.

\section{Conclusion}

We studied the properties of entanglement among scattering particles ($AB\rightarrow AB$, $CD\rightarrow CD$), based on the double scattering processes $e^+e^-\rightarrow\mu^+\mu^-$, wherein electron $A$ is initially entangled with electron $C$. Using the perturbative method, we studied the change in the mutual information in subsystems AC, BD, and AB. The results show that, for an initially polarized positron pair, the change in mutual information in subsystems A and B is proportional to the total cross section $\sigma_{tot}$ of double scattering processes, rather than the cross section of one scattering process. Particularly, the entanglement between subsystems A and C still changes proportionally to the total cross section, even they do not have direct interactions with each other during the entire scattering process. Our calculations show that the bipartite entanglement distributes among all pairs of subsystems after the scattering process, indicating some kind of entanglement sharing property in multipartite system.
Using the four-qubit filter operators given by Osterloh and Siewert, we calculated the entanglement monotones of the final state in the double-scattering process. The results show that, for the double-scattering process $e^+e^-\rightarrow\mu^+\mu^-$, the scattering final state is not a genuine entangled state, and the state of the four entangled particles resembles the W entangled state.

\section*{Acknowledgment}

The authors would like to thank Jin-Xuan Zhao and Yanbin Deng for useful discussions. This work is supported by the National Natural Science Foundation of China (Grants No. 11947046 and No. 11947086).


\begin{thebibliography}{100}

\bibitem{Takayanagi:2006} S. Ryu, T. Takayanagi, Holographic Derivation of Entanglement Entropy from AdS/CFT, Phys. Rev. Lett. \textbf{96}, 181602 (2006).

\bibitem{Pastawski:2015qua}
F. Pastawski, B. Yoshida, D. Harlow and J. Preskill, Holographic quantum erro-correcting codes: Toy models for the bulk/boundary correspondence, JHEP {\bf 1506}, 149 (2015).

\bibitem{Brown:2015lvg} 
  A. R. Brown, D. A. Roberts, L. Susskind, B. Swingle and Y. Zhao,
  Complexity, action, and black holes, Phys. Rev. D {\bf 93}, 086006 (2016).

\bibitem{Hsu:2012gk}
  T. C. L. Hsu, M. B. McDermott and M. Van Raamsdonk,
  Momentum-space entanglement for interacting fermions at finite density, JHEP, {\bf 1311}, 121 (2013).

\bibitem{Balasubramanian:2011wt}
  V. Balasubramanian, M. B. McDermott and M. Van Raamsdonk,
  Momentum-space entanglement and renormalization in quantum field theory, Phys. Rev. D {\bf 86}, 045014 (2012).

\bibitem{Park:2014hya}
  S. Seki, I. Y. Park and S. J. Sin,
  Variation of Entanglement Entropy in Scattering Process, Phys. Lett. B {\bf 743}, 147 (2015).

\bibitem{Peschanski:2016} R. Peschanski, S. Seki, Entanglement Entropy of Scattering Particles,
 Phys. Lett. B. \textbf{758}, 89 (2016).
 
\bibitem{Fan:2017hcd}
  J. Fan, Y. Deng and Y. C. Huang,
  Variation of entanglement entropy and mutual information in fermion-fermion scattering, Phys. Rev. D {\bf  95}, 065017 (2017).

\bibitem{Fan:2017mth} 
  J. Fan and X. Li,
  Relativistic effect of entanglement in fermion-fermion scattering,
Phys. Rev. D {\bf 97}, 016011 (2018).

\bibitem{Sampaio:2016} R. Faleiro, R. Pavão, H. Alexander, B. Hiller, A. Blin, M. Sampaio,   Momentum correlations of scattered particles in quantum field theory: one-loop entanglement generation, arXiv:1607.01715.
  
\bibitem{Semenoff:2016} D. Carney, L. Chaurette, G. Semenoff, Scattering with partial information, arXiv:1606.03103.

\bibitem{Ratzel:2016qhg}
  D. Rätzel, M. Wilkens and R. Menzel,
 Effect of polarization entanglement in photon-photon scattering,
 Phys. Rev. A {\bf 95}, 012101 (2017).

\bibitem{Araujo:2019mni} 
  J. B. Araujo, B. Hiller, I. G. da Paz, M. M. Ferreira, Jr., M. Sampaio and H. A. S. Costa, Measuring QED cross sections via entanglement, Phys. Rev. D {\bf 100}, 105018 (2019).

\bibitem{Bennett:1996gf} 
  C. H. Bennett, D. P. DiVincenzo, J. A. Smolin and W. K. Wootters,
  Mixed state entanglement and quantum error correction, Phys. Rev. A {\bf 54}, 3824 (1996).

\bibitem{Hill:1997pfa} 
  S. Hill and W. K. Wootters, Entanglement of a pair of quantum bits,
Phys. Rev. Lett. {\bf 78}, 5022 (1997).

\bibitem{Wootters:1997id} 
  W. K. Wootters, Entanglement of formation of an arbitrary state of two qubits, Phys. Rev. Lett. {\bf 80}, 2245 (1998).

\bibitem{Osterloh:2005}
A. Osterloh and J. Siewert, Constructing N-quit entanglement monotones from antilinear operators, Phys. Rev. A {\bf 72}, 012337.

\bibitem{Thomson}
M. Thomson,
Modern Particle Physics,
Cambridge University Press, N. Y. (2013).

\bibitem{Weinberg}
S. Weinberg,
The Quantum Theory of Fields I,
Cambridge University Press, N. Y. (1995).

\bibitem{Verstraete:2002}
F. Verstraete, J. Dehaene, B. De Moor, and H. Verschelde, Four qubits can be entangled in nine different ways, Phys. Rev. A {\bf 65}, 052112.

\bibitem{Osterloh:2006}
A. Osterloh, J. Siewert, Entanglement monotones and maximally entangled states in multipartite qubit systems, Int. J. Quant. Inf. 4, 531 (2006).

\end{thebibliography}
\end{document}